# Optimal Skeleton Network Reconfiguration considering Topological Characteristics and Transmission Path

Jin Lu
Department of Electrical and Computer Engineering
University of Houston
Houston, TX, USA
jlu28@uh.edu

Xingpeng Li
*Member, IEEE*
Department of Electrical and Computer Engineering
University of Houston
Houston, TX, USA
xli82@uh.edu

*Abstract*—Power system restoration can be divided into three stages: black-start, network reconfiguration and load restoration. A skeleton-network should be restored in the second stage to prepare for the subsequent large-scale system-wide load pickup. This paper proposes a novel integrated skeleton-network reconfiguration (ISNR) model which considers both the topological characteristics of the network and the transmission path energization constraints. A novel topological characteristics-based skeleton-network quality index (TCSNQI), an index based on the network importance and distance, is proposed to evaluate the quality of the skeleton-network. The proposed ISNR model can attain both the target network and the associated restoration sequence for that network. The attained target network reaches a certain quality level, and it requires the least restoration time based on the attained sequence of ordered switching actions. The proposed ISNR model is formulated as a mixed-integer linear programming (MILP) problem. Numerical simulations on the New England 39-bus test system demonstrate the performance of the proposed ISNR model.

*Index Terms*—Black start, Mixed-integer linear programming, Network reconfiguration, Power system restoration, Skeleton network restoration.

## Nomenclature

*Sets*
$B$ — Set of all buses in the power system.
$B^C$ — Set of critical buses in critical transmission paths.
$B^{NC}$ — Set of all buses except the critical buses.
$L$ — Set of all branches in the power system.
$L^C$ — Set of critical branches in critical transmission paths.
$L_i^N$ — Set of all branches connected to bus $i$.
$T$ — Set of time steps.

*Parameters*
$d_{ij}$ — The total number of branches in the shortest path between bus $i$ and bus $j$.
$d_{min,ij}$ — The total number of branches in the shortest path between bus $i$ and bus $j$ after node $i$ contraction.
$l_i$ — The average distance of the contracted network.
$N$ — The network quality index requirement of the target skeleton-network.
$n_i$ — The total bus number of the network after the contraction of bus $i$.
$u_{i,t}^C$ — Indicator of energization status of critical bus $i$ at time $t$ from the black-start solution.
$u_{L_{ij},t}^C$ — Indicator of energization status of critical branch $L_{ij}$ at time $t$ in the black-start solution.
$\alpha_i$ — The bus importance degree of bus $i$.
$\beta$ — The weight coefficient of network importance term.
$\gamma$ — The weight coefficient of network distance term.

*Variables*
$d_t$ — The total distance between buses in the restored network at time $t$.
$u_t^S$ — Skeleton-network quality status at time $t$.
$u_{i,t}$ — Indicator of energization status of bus $i$ at time $t$.
$u_{L_{ij},t}$ — Indicator of energization status of branch $L_{ij}$ at time $t$.
$v_t^S$ — Indicator of skeleton-network quality status change at time $t$.
$v_{i,t}$ — Indicator of energization status change of bus $i$ at time $t$.
$y_{i,j,t}^1$ — First ancillary variable for buses $i$ and $j$ at time $t$.
$y_{i,j,t}^2$ — Second ancillary variable for buses $i$ and $j$ at time $t$.
$\eta_t^S$ — The skeleton-network quality index of the skeleton-network at time $t$.

## I. Introduction

THe power system plays an important role in modern society. In [1], the global electricity demand is forecasted to keep increasing to 2040 in the Stated Policies Scenario. With the development and expansion of the power system, the grid reliability and resilience are of vital significance. However, a number of large-scale blackouts still occur around the world. Many different factors can cause a blackout, such as natural disasters and equipment failure [2]. An effective power restoration strategy is one of the keys to achieving a resilient bulk power system against outages.

The aim of the restoration strategy is to restore the maximum loads in the shortest time under the power system safety constraints. Power system restoration is a multi-objective, multi-stage, multi-variable and multi-constraint optimization problem [3]. Traditionally, the restoration process is separated into three stages: black-start, network reconfiguration and load restoration. In the black-start stage, the non-black-start (NBS) generators will be recovered by utilizing the black-start generator. In the network reconfiguration stage, a stable backbone network with majority substations is established to enable load restoration in the next stage. In the load restoration stage, large-scale load restoration can be carried out [4]. Reference [5] proposes a general restoration strategy consisting of six restoration steps called generic restoration milestones. In [6], a two-stage adaptive restoration decision support system is proposed. The optimal planning function determines the operation sequence and the optimal real-time function adjusts values of generation and load pickup. In addition to the research on the entire restoration process, many studies focus on a single stage of the restoration. For instance, [7]-[10] investigate the black-start stage to find the generator start-up sequence (GSUS); [11]-[13] explore the network reconfiguration stage while [14]-[16] examine the load pickup stage. In this paper, we mainly focus on the network reconfiguration stage.

In the second stage of power system restoration, the reconstruction of the backbone network can be divided into two steps: (a) determine the target skeleton-network and (b) identify the optimal restoration sequence of the network determined in the previous step. A number of prior works have been conducted to determine the target network. In [17], it introduces node importance degree to evaluate the importance for every single bus. The network reconfiguration efficiency is defined based on the node importance degree and the average clustering coefficient of the network. The target skeleton-network with maximum network reconfiguration efficiency is obtained using discrete particle swarm optimization. In [18], it defines a new network index considering both node importance and line charging reactive power, the target skeleton-network is obtained by genetic algorithm. In [19], it defines a new network index as a linear combination of four different network indexes considering the cost of blackout in market environments. In [20], seven types of network indexes are synthesized using criteria importance through intercriteria correlation to comprehensively evaluate the node importance. In [21], it identifies the network restoration sequence based on the determined target network. It proposes a comprehensive skeleton-network restoration strategy that includes three models for GSUS, skeleton-network and load pickup respectively. A traditional skeleton-network model, namely the transmission line restoration (TLR) model, is used in [21] and it can attain the restoration sequence based on the target network obtained by the method in [17] and the GSUS solution achieved from the GSUS model. However, the skeleton-network obtained by [14] does not consider its restoration performance such as the total restoration time of the skeleton-network.

In summary, a method to determine the skeleton-network considering both the quality and restoration of the network is lacking in the literature. This paper bridges this gap and presents an integrated skeleton-network reconfiguration (ISNR) model which can determine both the target network and the corresponding restoration sequence. Two topology characteristics are introduced to evaluate the network quality: the network importance and the network distance. Besides, the transmission path energization constraints are included to optimize the restoration time of the target network and provide a more practical restoration sequence. The proposed ISNR model is an MILP problem that can be efficiently solved to obtain the optimal solution.

The rest of this work is structured as follows. Section II explains a general network reconfiguration restoration strategy and the proposed ISNR model-based restoration strategy. Section III presents the formulation of the proposed ISNR model. Section IV presents the numerical simulation results. The conclusions are drawn in Section V.

## II. RESTORATION STRATEGIES OF THE NETWORK RECONFIGURATION STAGE

The network reconfiguration stage is to reconstruct an efficient and reliable skeleton-network which is the basis of the large-scale load restoration. In the black-start stage, some buses and branches are restored to deliver the cranking power to NBS generators. These critical buses and branches are necessary for restoration and should be included in the target skeleton network. In other words, the restoration strategy of skeleton-network should take the black-start solution as an input to determine the target skeleton-network. After the skeleton-network and the corresponding restoration sequence are determined, load pickup actions should be taken to eliminate the voltage violation problem during the energization of the obtained skeleton-network. A general five-step restoration strategy is introduced to demonstrate the decision procedures of the network reconfiguration.

### A. A general five-step restoration strategy

In general, the restoration strategy of network reconfiguration can be divided into five steps. Each step can be considered as a tractable problem. The five-step restoration strategy is described below: (a) find the optimal generator start up sequence and critical transmission paths to crank the generators; (b) determine a well-performed skeleton network; (c) identify the restoration sequence of the skeleton network found by step b; (d) determine the load pickup during the restoration to eliminate the voltage violation; (e) check whether all safety constraints are satisfied; if the current restoration solution fails the check, it will go back to step a and specific constraints will be added based on the violations. The general skeleton-network restoration strategy is illustrated in Fig. 1.

### B. The proposed ISNR model-based restoration strategy

The proposed ISNR model can simultaneously determine the skeleton-network and the corresponding restoration sequence. The detailed formulation of proposed ISNR model is presented in Section II. The critical branches and buses identified in the black-start stage will be the input data of the proposed ISNR model. Besides, the restoration sequence determined by the proposed ISNR model can be used and modified in the load pickup step. This indicates the proposed ISNR model can realize the functions of both step b and step c. The skeleton-network and the restoration sequence are co-optimized to achieve an efficient and fast recoverable skeleton-network restoration solution. The restoration strategy with the proposed ISNR model is illustrated in Fig. 2. This paper focuses on step b and step c and examines the proposed ISNR model. Other steps for the restoration strategy are not discussed in this paper.

Both the proposed ISNR model and the benchmark TLR model (corresponding to a sequential implementation of step b and step c) use the black-start solution obtained by the algorithm in [21] as inputs.

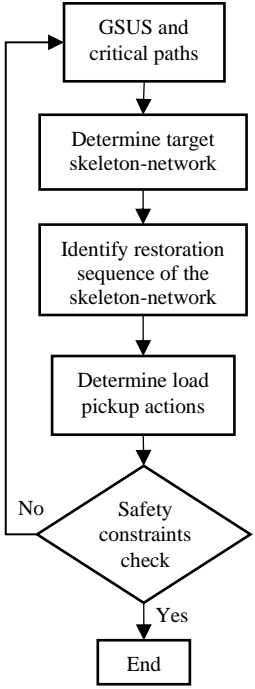

Fig. 1. Flowchart of network reconfiguration strategy.

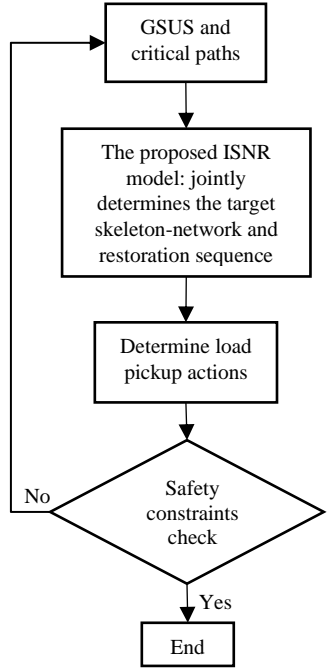

Fig. 2. The strategy with the proposed ISNR model.

## III. FORMULATION OF THE PROPOSED ISNR MODEL

### A. Skeleton-network formulation

To achieve a well-performed skeleton-network that can be restored in the least time, both the skeleton-network quality and the restoration time should be considered in the model. A topology characteristics-based skeleton-network quality index (TCSNQI) is proposed in this paper to evaluate the quality of the network at each time step. The network which achieves a quality index larger than a specific value will be regarded as a qualified skeleton-network. A qualified skeleton-network should be restored as soon as possible. Thus, as expressed in (1), the objective function is to minimize the total time that a skeleton-network does not satisfy the skeleton quality requirement.

$$\min \sum_{t \in T}(1 - u_t^S) \quad (1)$$

where $u_t^S$ is the binary indicator of the skeleton-network quality. When $u_t^S$ is 1, the restored network at time $t$ satisfies the quality requirement; otherwise, the restored network fails to meet the requirement at time $t$.

In (2), the skeleton-network will keep the qualified status after it meets the quality requirement. In (3), the indicator of quality status change $v_t^S$ can be 1 only when the skeleton-network meets the quality requirement at time $t$.

$$u_{t-1}^S \leq u_t^S, \quad \forall t \in T \quad (2)$$

$$v_t^S = u_t^S - u_{t-1}^S, \quad \forall t \in T \quad (3)$$

The TCSNQI proposed to evaluate the quality of the network at each time step is defined as follows.

$$\eta_t^S = \beta \sum_{i \in B} \alpha_i u_{i,t} + \gamma d_t, \quad \forall t \in T \quad (4)$$

$$\alpha_i = \frac{1}{n_i \cdot l_i} \quad (5)$$

$$l_i = \frac{\sum_{i,j \in V_i} d_{min,ij}}{n_i(n_i - 1)/2} \quad (6)$$

$$d_t = d_{t-1} + \sum_{i \in B} \sum_{j \in B, j \neq i} d_{ij} v_{i,t} u_{j,t-1} + \sum_{i \in B} \sum_{j \in B, j \neq i} d_{ij} v_{i,t} v_{j,t}/2, \quad (7)$$
$$\forall t \in T$$

In (4), the proposed TCSNQI is the sum of a weighted network importance term and a weighted network distance term. The network importance term is the total bus importance degree of the buses restored by time $t$, where $\alpha_i$ denotes the bus importance degree of bus $i$, and $u_{i,t}$ represents the binary variable of restoration status for bus $i$ at time $t$. The bus importance degree proposed in [17] is defined by (5) and (6). The network distance term is defined by (7). The network distance is the total distance between the restored buses at time $t$. A larger network distance means the restored skeleton-network can cover larger area. The distance between the buses is calculated for the fully restored power system, thus the skeleton-network can cover a large area from the perspective of the entire system. The network distance is the sum of the previous network distance and the distance caused by newly restored buses at time $t$. The total distance between the newly restored buses and previously restored buses is calculated by the second term in (7). The total distance between the newly restored buses is calculated by the third term in (7). $d_{i,j}$ denotes the distance between bus $i$ and bus $j$, which is the total number of branches in the shortest path between the two buses.

When the skeleton-network meets the quality requirement at time $t$, the proposed TCSNQI should be larger than a prespecified number $N$. It can be described by constraint (8).

$$\eta_t^S \geq N \cdot v_t^S, \quad \forall t \in T \quad (8)$$

### B. Transmission path energization constraints

The transmission path energization constraints are shown as (9)-(16).

$$u_{L_{ij},t} \leq u_{i,t}, \quad \forall L_{ij} \in L, \forall t \in T \quad (9)$$

$$u_{L_{ij},t} \leq u_{j,t}, \quad \forall L_{ij} \in L, \forall t \in T \quad (10)$$

$$u_{L_{ij},t+1} \leq u_{i,t} + u_{j,t}, \quad \forall L_{ij} \in L, \forall t \in T \quad (11)$$

$$u_{i,t} \leq \sum_{L_{ij} \in L_i^N} u_{L_{ij},t}, \quad \forall i \in B^{NC}, \forall t \in T \quad (12)$$

$$u_{L_{ij},t} \leq u_{L_{ij},t+1}, \quad \forall L_{ij} \in L, \forall t \in T \quad (13)$$

$$v_{i,t} = u_{i,t} - u_{i,t-1}, \quad \forall i \in B, \forall t \in T \quad (14)$$

$$u_{L_{ij},t} = u_{L_{ij},t}^C, \quad \forall L_{ij} \in L^C, \forall t \in T \quad (15)$$

$$u_{i,t} = u_{i,t}^C, \quad \forall i \in B^C, \forall t \in T \quad (16)$$

If a branch is energized, its terminal buses can be energized by (9)-(10). In (11), a branch can be energized at the next time step if at least one of the terminal buses is energized. (12) indicates a bus can be energized when at least one of its connected branches is restored. Constraint (13) ensures the branch will remain energized after it is restored. Equation (15) and (16) guarantee the energization of critical branches and buses identified in the black start stage. $u_{L_{ij},t}^C$ and $u_{i,t}^C$ are the energization indicators of the critical branches and buses at time $t$ in the black-start stage respectively.

*C. Linearization*

The product terms between two binary variables in (7) will lead to non-linearity. To address this issue, two ancillary variables $y_{i,j,t}^1$ and $y_{i,j,t}^2$ that are defined in (17) are introduced to linearize (7). First, (14) is substituted into (7) to obtain (18) and then, (17) is substituted into (18) to obtain the linearized expression (19).

$$\begin{cases} y_{i,j,t}^1 = u_{i,t} \cdot u_{j,t}, & \forall t \in T \\ y_{i,j,t}^2 = u_{i,t-1} \cdot u_{j,t}, & \forall t \in T \end{cases} \quad (17)$$

$$d_t = d_{t-1} + \sum_{i \in B} \sum_{j \in B, j \neq i} d_{i,j}(u_{i,t} \cdot u_{j,t} - u_{i,t-1} \cdot u_{j,t}) \\ + \sum_{i \in B} \sum_{j \in B, j \neq i} d_{i,j}(u_{i,t} \cdot u_{j,t} \\ - u_{i,t-1} \cdot u_{j,t} - u_{i,t} \cdot u_{j,t-1} \\ + u_{i,t-1} \cdot u_{j,t-1})/2, \forall t \in T \quad (18)$$

$$d_t = d_{t-1} + \sum_{i \in B} \sum_{j \in B, j \neq i} d_{i,j}(1.5 y_{i,j,t}^1 \\ - 1.5 y_{i,j,t}^2 - 0.5 y_{j,i,t}^2 \\ + 0.5 y_{i,j,t-1}^1), \forall t \in T \quad (19)$$

Since the definition of ancillary variables $y_{i,j,t}^1$ and $y_{i,j,t}^2$ (17) still involves non-linearity, a set of linear inequations (20)-(21) will be used to replace (17) in the proposed ISNR model.

$$\begin{cases} y_{i,j,t}^1 \geq u_{i,t} + u_{j,t} - 1, & \forall t \in T \\ y_{i,j,t}^1 \leq u_{i,t}, & \forall t \in T \\ y_{i,j,t}^1 \leq u_{j,t}, \forall t \in T \end{cases} \quad (20)$$

$$\begin{cases} y_{i,j,t}^2 \geq u_{i,t-1} + u_{j,t} - 1, & \forall t \in T \\ y_{i,j,t}^2 \leq u_{i,t-1}, & \forall t \in T \\ y_{i,j,t}^2 \leq u_{j,t}, & \forall t \in T \end{cases} \quad (21)$$

To summarize, the proposed ISNR model consists of (1)-(6), (8)-(16), (19)-(21)(21).

## IV. CASE STUDIES

The proposed ISNR model is tested using the New England 39-bus system shown in Fig. 3. The time interval is 5 minutes. The weight coefficient $\beta$ is set to 150, and $\gamma$ is set to 0.7. The least requirement $N$ for skeleton-network quality index is set to 1,500. The test case is solved using solver Gurobi in Python. The simulation is performed on a desktop computer with Intel-i7 3.2 GHz CPU and 16 GB RAM.

The skeleton-network meets the network quality requirement at 35 minutes according to the optimal objective value. A quality requirement which is easier to achieve is applied, since the restoration can be delayed due to the safety constraints and other reasons. It can be noticed that some critical buses and branches are not restored when the network is qualified. The skeleton-network determined by the proposed ISNR model includes 32 skeleton buses and 31 skeleton branches. The optimal skeleton buses are listed in Table I. It can be noticed that bus 9 is added to the skeleton-network besides the critical buses found by the GSUS model in [21].

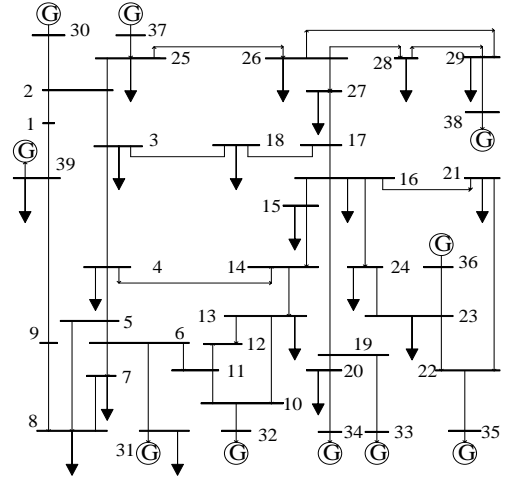

Fig. 3. The New England 39-bus system architecture [22].

TABLE I
SKELETON NETWORK BUSES OF THE ISNR MODEL

| Bus Number | Bus Start Time (Minute) | Bus Importance Degree | Bus Number | Bus Start Time (Minute) | Bus Importance Degree |
|---|---|---|---|---|---|
| 1 | 25 | 0.812 | 22 | 50 | 0.845 |
| 2 | 20 | 0.903 | 23 | 50 | 0.845 |
| 3 | 25 | 0.907 | 24 | 45 | 0.823 |
| 4 | 30 | 0.916 | 25 | 25 | 0.861 |
| 5 | 35 | 0.883 | 26 | 30 | 0.894 |
| 6 | 40 | 0.883 | 29 | 35 | 0.840 |
| 9 | 35 | 0.812 | 30 | 15 | 0.770 |
| 10 | 45 | 0.845 | 31 | 45 | 0.776 |
| 13 | 40 | 0.851 | 32 | 50 | 0.778 |
| 14 | 35 | 0.911 | 33 | 50 | 0.773 |
| 16 | 40 | 1 | 34 | 65 | 0.781 |
| 17 | 35 | 0.904 | 35 | 55 | 0.780 |
| 18 | 30 | 0.846 | 36 | 55 | 0.780 |
| 19 | 45 | 0.839 | 37 | 30 | 0.773 |
| 20 | 60 | 0.809 | 38 | 40 | 0.781 |
| 21 | 45 | 0.823 | 39 | 30 | 0.812 |

The skeleton branches determined by the proposed ISNR model are shown in Table II. Branch 9-39 is restored at $t$=35 minutes besides the critical branches. Bus 9 can be energized due to the energization of branch 9-39. It lays the foundation for restoring load bus 8 and load bus 7.

TABLE II
SKELETON NETWORK BRANCHES OF ISNR MODEL

| Branch (from bus – to bus) | Start Time (Minute) | Branch (from bus – to bus) | Start Time (Minute) | Branch (from bus – to bus) | Start Time (Minute) |
|---|---|---|---|---|---|
| 1-2  | 25 | 10-32 | 50 | 23-36 | 55 |
| 1-39 | 30 | 13-14 | 40 | 25-26 | 30 |
| 2-3  | 25 | 16-17 | 40 | 25-37 | 30 |
| 2-25 | 25 | 16-19 | 45 | 26-29 | 35 |
| 2-30 | 20 | 16-21 | 45 | 29-38 | 40 |
| 3-4  | 30 | 16-24 | 45 | | |
| 3-18 | 30 | 17-18 | 35 | | |
| 4-5  | 35 | 19-20 | 60 | | |
| 4-14 | 35 | 19-33 | 50 | | |
| 5-6  | 40 | 20-34 | 65 | | |
| 6-31 | 45 | 21-22 | 50 | | |
| 9-39 | 35 | 22-35 | 55 | | |
| 10-13 | 45 | 23-24 | 50 | | |

The simulation results are compared with the TLR model used in [21]. The method in [18] found a skeleton network with 32 skeleton buses and 32 skeleton branches. Although the two skeleton-networks have the same number of skeleton buses, they do not share the same skeleton buses. The TLR model selects bus 11 as the skeleton bus instead of bus 9 that is selected by the proposed ISNR model. The total bus importance degree of skeleton-network is 26.8 for the proposed ISNR model which is the same with the TLR model. Fig. 4-6 shows the plots of network importance, network distance and network quality index vs. time respectively. The proposed ISNR model has larger values for three indexes in the restoration period from 35 minutes to 50 minutes, which demonstrates the proposed ISNR model outperforms the traditional TLR model. This is because the proposed ISNR model integrates the transmission path and topology characteristics that would lead to a more efficient and faster recoverable skeleton network.

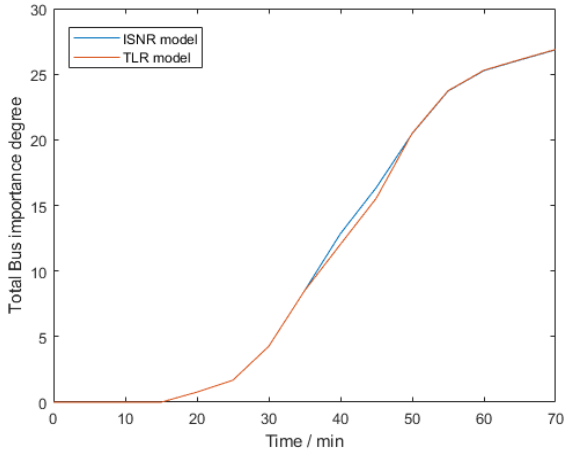

Fig. 4. Comparison of total bus importance degree between the proposed ISNR and the traditional TLR models.

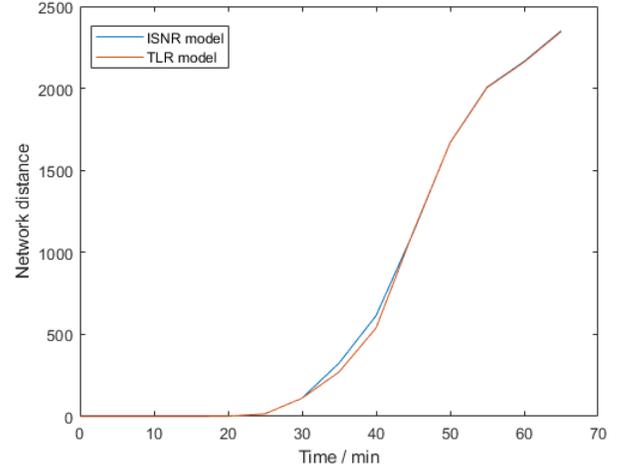

Fig. 5. Comparison of the network distance between the proposed ISNR and the traditional TLR models.

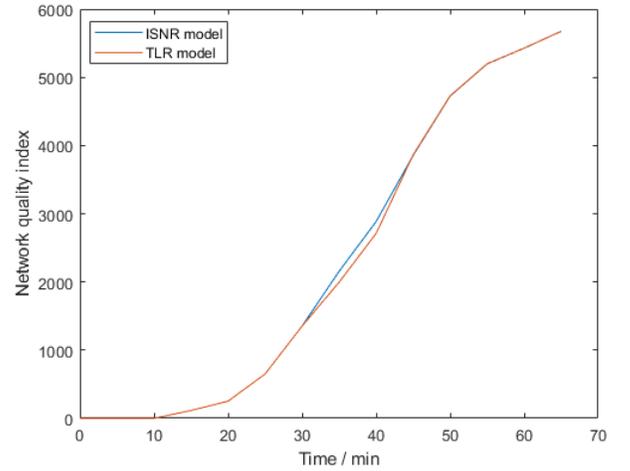

Fig. 6. Comparison of the network quality index between the proposed ISNR and the traditional TLR models.

## V. CONCLUSION

In this paper, a new skeleton-network reconfiguration model, namely the ISNR model is proposed. The proposed ISNR model aims to obtain a well-performed and fast recoverable skeleton-network as well as the associated restoration sequence of it. The proposed TCSNQI is used to evaluate the quality of the network considering both the bus importance and the total distance between the buses in the network. The proposed ISNR model will determine a target network that has the least restoration time while meeting the specified quality index requirement. The simulation results on the New England 39-bus system demonstrate the proposed method can obtain an efficient skeleton-network during the restoration process. The restoration sequence solution may be modified since the load pickup actions during the skeleton-network reconfiguration stage are not determined and the safety constraints such as voltage limitation may be violated. A comprehensive restoration strategy including the load pickup function and safety constraints check function as illustrated in Fig. 2 need to be investigated in future work.